\documentclass[sigconf]{acmart}
\copyrightyear{2025}
\acmYear{2025}
\setcopyright{acmlicensed}\acmConference[SIGIR '25]{Proceedings of the 48th International ACM SIGIR Conference on Research and Development in Information Retrieval}{July 13--18, 2025}{Padua, Italy}
\acmBooktitle{Proceedings of the 48th International ACM SIGIR Conference on Research and Development in Information Retrieval (SIGIR '25), July 13--18, 2025, Padua, Italy}
\acmDOI{10.1145/3726302.3731968}
\acmISBN{979-8-4007-1592-1/2025/07}

\AtBeginDocument{%
  \providecommand\BibTeX{{%
    \normalfont B\kern-0.5em{\scshape i\kern-0.25em b}\kern-0.8em\TeX}}}
\usepackage[linesnumbered,ruled,vlined]{algorithm2e}
\usepackage{multirow}
\usepackage{stfloats}
\usepackage{balance}
\begin{document}
\title{Pyramid Mixer: Multi-dimensional Multi-period Interest Modeling for Sequential Recommendation}

\author{Zhen Gong}
\email{gongzhen.666@bytedance.com}
\affiliation{
  \institution{Bytedance}
  \city{Shanghai}
  \country{China}
}

\author{Zhifang Fan}
\email{fanzhifangfzf@gmail.com}
\affiliation{
  \institution{Bytedance}
  \city{Shanghai}
  \country{China}
}

\author{Hui Lu}
\email{luhui.xx@bytedance.com}
\affiliation{
  \institution{Bytedance}
  \city{Hangzhou}
  \country{China}
}

\author{Qiwei Chen$^{*}$}
\email{chenqiwei05@gmail.com}
\affiliation{
  \institution{Bytedance}
  \city{Shanghai}
  \country{China}
}

\author{Chenbin Zhang}
\email{aleczhang13@gmail.com}
\affiliation{
  \institution{Bytedance}
  \city{Beijing}
  \country{China}
}

\author{Lin Guan}
\email{guanlin.13@gmail.com}
\affiliation{
  \institution{Bytedance}
  \city{Beijing}
  \country{China}
}

\author{Yuchao Zheng}
\email{zhengyuchao.yc@bytedance.com}
\affiliation{
  \institution{Bytedance}
  \city{Hangzhou}
  \country{China}
}

\author{Feng Zhang}
\email{feng.zhang@bytedance.com}
\affiliation{
  \institution{Bytedance}
  \city{Shanghai}
  \country{China}
}

\author{Xiao Yang}
\email{wuqi.shaw@bytedance.com}
\affiliation{
  \institution{Bytedance}
  \city{Beijing}
  \country{China}
}

\author{Zuotao Liu}
\email{michael.liu@bytedance.com}
\affiliation{
  \institution{Bytedance}
  \city{Singapore}
  \country{Singapore}
}
\renewcommand{\shortauthors}{Zhen Gong et al.}


\thanks{$^*$Qiwei Chen is the corresponding author.}

\begin{abstract}
Sequential recommendation, a critical task in recommendation systems, predicts the next user action based on the understanding of the user's historical behaviors. Conventional studies mainly focus on cross-behavior modeling with self-attention based methods while neglecting comprehensive user interest modeling for more dimensions. In this study, we propose a novel sequential recommendation model, Pyramid Mixer, which leverages the MLP-Mixer architecture to achieve efficient and complete modeling of user interests. Our method learns comprehensive user interests via cross-behavior and cross-feature user sequence modeling. The mixer layers are stacked in a pyramid way for cross-period user temporal interest learning. Through extensive offline and online experiments, we demonstrate the effectiveness and efficiency of our method, and we obtain a +0.106\% improvement in user stay duration and a +0.0113\% increase in user active days in the online A/B test. The Pyramid Mixer has been successfully deployed on the industrial platform, demonstrating its scalability and impact in real-world applications.
\end{abstract}


\maketitle
\section{Introduction}
Recommendation systems have been widely deployed in online platforms for user interest modeling. User browsing and interaction behaviors provide dynamic and rich information of user preferences. User behavior sequences also record annual, seasonal, and daily user interests. Modeling historical user behaviors and item features within and between periods is essential for these systems to accurately understand and predict personalized user interests.
  
\begin{figure}[t]\vspace{-1.4\baselineskip}\small\centering
  \includegraphics[width=1\linewidth]{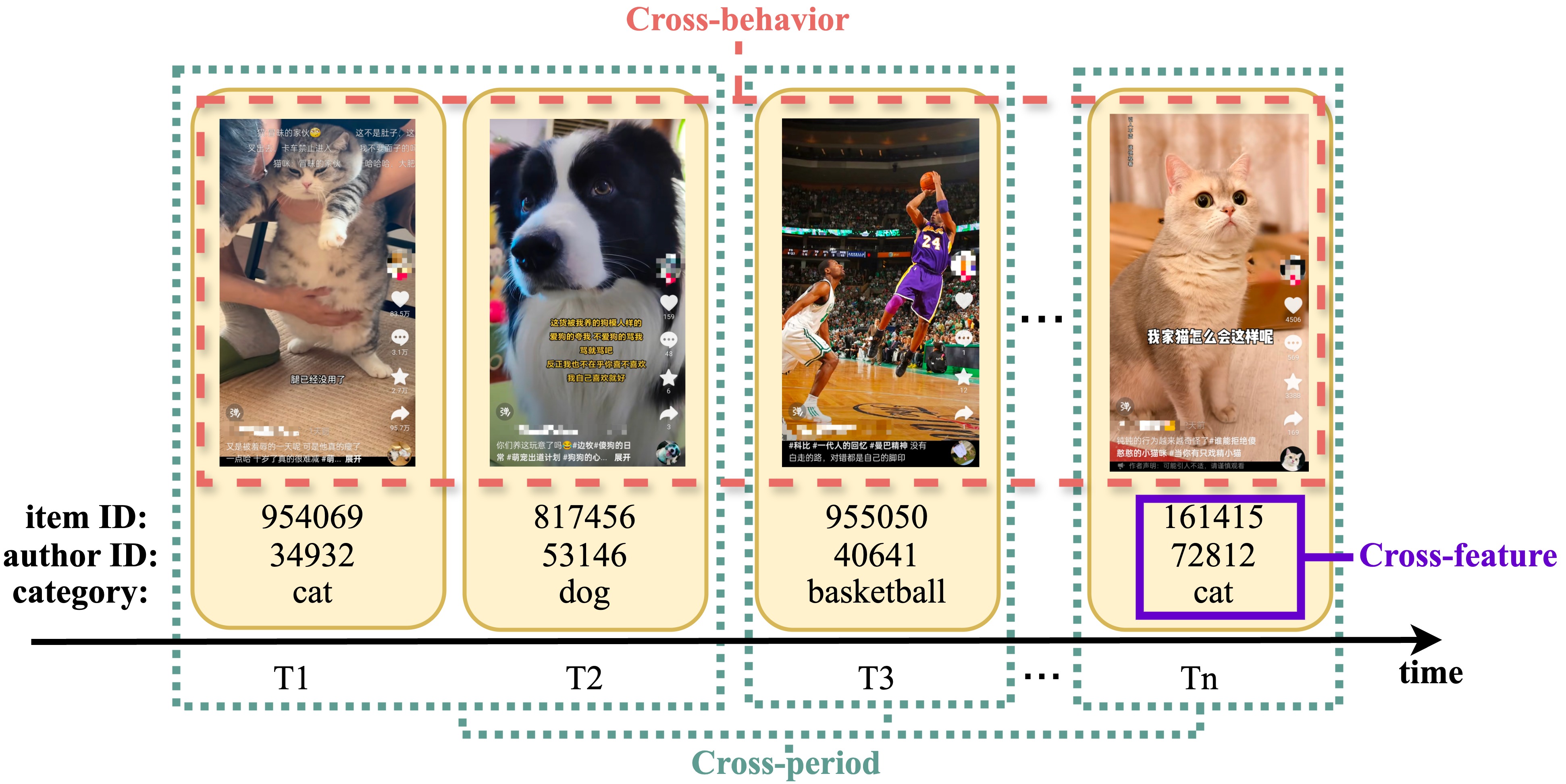}
  \vspace{-2\baselineskip}
  \caption{Cross-behavior, feature and period modeling.}
  \vspace{-2\baselineskip}
  \label{motivation}
\end{figure}

Most existing studies focus on capturing the hidden interests in sequences via self-attention mechanism~\cite{kang2018self,wu2019neural,fan2021lighter}. They mainly focus on modeling the relationship between behaviors while neglecting the feature crossing of IDs and side information of each behavior. These self-attention based methods largely lack a comprehensive cross-behavior and cross-feature mixing of user interests. Given that the computation complexity of self-attention is $O(n^2)$, these methods are hard to be applied in large industrial scenarios under strict computational constraints. However, the all-MLP architecture, such as MLP-Mixer~\cite{Mlp-mixer} and ResMLP~\cite{touvron2022resmlp}, may be a key to the problem. The all-MLP architecture demonstrates competitive performance and attracts the attention of both industry and academia due to its efficiency and simplicity. As for recommendation systems, there are also studies about recommendation models with pure MLPs~\cite{fmlp-rec,mlp4rec,lee2021moi}. MLP-Mixer related studies are notable and inspiring given their architectural simplicity and linear computational complexity. Besides, many research works employ the pyramid or multi-scale feature fusion scheme to achieve multiple receptive fields. 
These pyramid multi-scale fusion methods have the potential to model user temporal interests at different time cycles.

In this paper, we propose the Pyramid Mixer model for comprehensive cross-behavior and cross-feature fusion of user behaviors. Pyramid Mixer leverages mixers to capture intricate correlations among user behaviors and item features in a parallel way. These mixers are hierarchically organized in a pyramid structure to effectively explore user interests across multiple scales. Furthermore, we integrate low-rank decomposition to enhance the computational efficiency of the Pyramid Mixer.

The main contributions of our work are summarized as follows:
\begin{itemize}
    \item We propose a simple MLP architecture with cross-behavior and cross-feature MLP mixers to capture comprehensive user interests among various behaviors and features of user behavior sequences.
    \item To enable our recommendation system to perceive multi-scale/period user preferences, the representations of mixers are hierarchically scaled from short-term to long-term interests in a pyramid user interest modeling scheme.
    \item Through online and offline experiments, we demonstrate the effectiveness of our model. We obtain a +0.106\% improvement in user stay duration and a +0.0113\% improvement on user active days. And the offline experiments also show it also achieves competitive performance.
\end{itemize}

\section{Related Work}
\textbf{Sequential Recommendation.} Sequential recommendation models aim to predict the next item that users are likely to interact with based on their historical behaviors. SASRec~\cite{kang2018self} leverages self-attention mechanisms to effectively capture user behavior patterns. GRU4Rec~\cite{hidasi2015session} utilizes Gated Recurrent Units~(GRUs) to model user behavior dynamics and uncover complex temporal dependencies. 

\textbf{All-MLP Architectures.} MLP-Mixer~\cite{Mlp-mixer} proposes a model that uses MLP in place of traditional attention-based methods to ensure efficiency. 
As for the recommendation research field, FMLP-Rec~\cite{fmlp-rec} is an all-MLP model with learnable filters to attenuate the noise information and maintain lower time complexity. MLP4Rec~\cite{mlp4rec} develops a tri-directional fusion scheme to coherently capture sequential, cross-channel, and cross-feature correlations, but its cascading structure leads to low efficiency.


\begin{figure*}[t]\small
  \vspace{-1\baselineskip}
  \centering
  \includegraphics[width=1\linewidth]{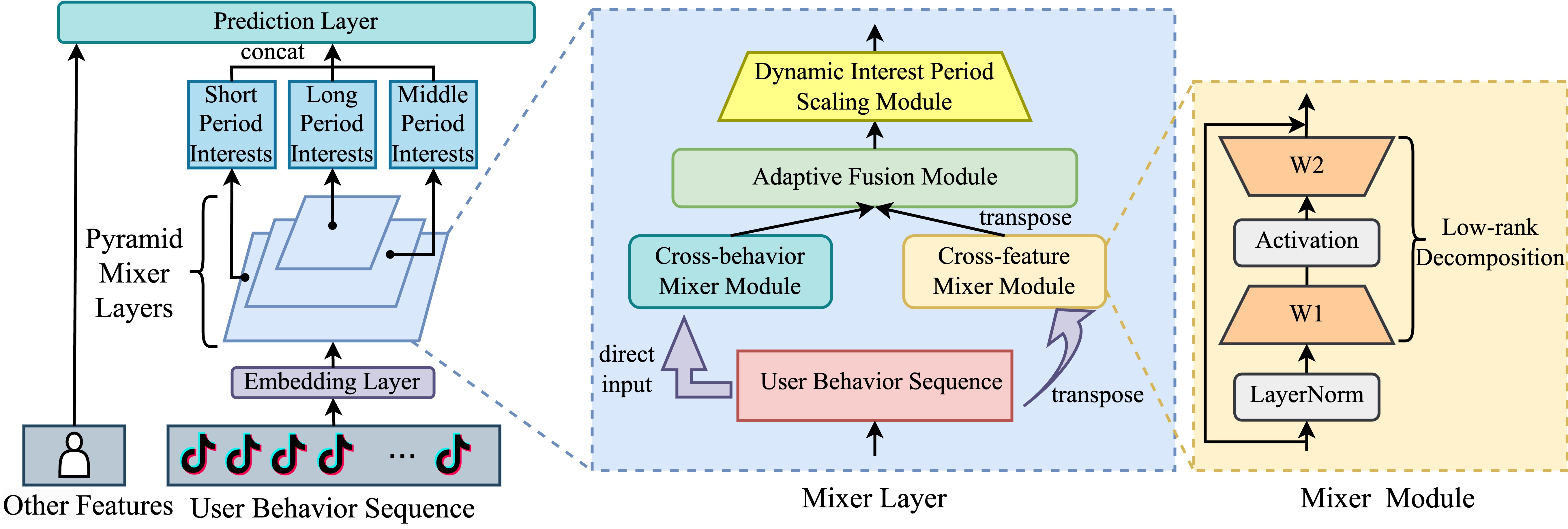}
  \vspace{-2\baselineskip}
  \caption{The architecture of Pyramid Mixer model.}
  \vspace{-1.5\baselineskip}
  \label{model}
\end{figure*}



\section{Pyramid Mixer}
\label{Pyramid Mixer}
Figure~\ref{model} illustrates the architecture of the proposed Pyramid Mixer model. The model adopts a pyramid-like structure, composed of mixer layers processing user sequence embedding at multiple scales. Within each mixer layer, cross-behavior and cross-feature mixer modules are employed to integrate user behaviors and item features, respectively. An adaptive fusion module is then utilized to seamlessly combine the outputs of these two mixer modules. The dynamic interest period scaling module further compresses the user behavior sequence into representations of varying scales. By stacking mixer layers hierarchically, the pyramid network constructs both short-period and long-period user interest feature maps, which are ultimately leveraged for prediction.

\textbf{Mixer Layers.} There are cross-behavior and cross-feature mixers in the mixer layer:
\begin{itemize}
    \item Cross-behavior Mixer Module: User sequences are directly input to the cross-behavior mixer to obtain user behavior representations and user interests with the feature interaction of different items in the same sequence. 
    \item Cross-feature Mixer Module: The cross-feature mixer outputs a comprehensive representation of each behavior. User sequences are first transposed and then fed to the mixer. The cross-feature mixer can mix features of each behavior, such as item ID, user actions~(e.g., playtime and likes), and other side information~(e.g., author ID and categories of the video).
\end{itemize}

Both cross-behavior and cross-feature mixer modules are implemented with pure MLPs. Each module has two fully connected layers and an activation function. And it could be formulated as: 
\begin{equation}\small
Y_i=X_i+W_2\sigma(W_1LayerNorm(X_i)),\\ i\in \{1,2,...,N_u\}
\end{equation}
where $X$ is the inputed sequence embedding, $Y$ is the output of the mixer module, $\sigma$ is the activation function~(Swish~\cite{ramachandran2017searching} or GELU~\cite{hendrycks2016gaussian}), and $W_1$ as well as $W_2$ are weight matrices. 

Since the over-parameterization problem is common in sequential recommendation tasks~\cite{fan2021lighter}, we apply a low-rank decomposition trick to weight matrices in order to reduce the computational cost and make our mixer layer more resilient to the over-parameterization problem. Specifically, the user behavior sequences are projected to a small number of latent interests with the first weight matrix $W_1$. After being compressed to lower dimensions, the latent interests are then recovered to the original sequence length for residual connections. These weight matrices are formulated as $W_1\in \mathbb{R}^{D\times D'}$ and $W_1\in \mathbb{R}^{D'\times D}$. The two hidden layer widths are unequal, and their relationship is $D'<D$. Decreasing the output width of $W_1$ and input width of $W_2$ involves fewer parameters and reduces computational overhead. 

We use an adaptive fusion module to combine the output of cross-behavior and cross-feature mixer modules. The adaptive fusion module generates a personalized weight $\alpha$ to adjust the trade-off between cross-behavior and cross-feature representations. The adaptive fusion module is a self-gate network implemented with a linear layer, and its input is the user behavior sequence.
\begin{equation}\small\begin{split}
\alpha_i &= \Sigma(Linear(X_i))\,,\\
Z_i &= \alpha_iY_i^{behavior}+(1-\alpha_i)Y_i^{feature}\,,
\end{split}\end{equation}
where $\alpha_i$ is the fusion weight, $\Sigma$ is a Sigmoid function, $Z_i$ is the output of adaptive fusion module, $Y_i^{behavior}$ is the output of cross-behavior mixer module and $Y_i^{feature}$ is the output of cross-feature mixer module.

Lastly, the dynamic interest period scaling module maps the sequence embedding to different scales for cross-period modeling.

\textbf{Pyramid Multi-Scale Cross-period Modeling.} The Pyramid Mixer model is built as a hierarchical pyramid encoder to discover temporal correlations across different sequence scales and capture multi-periodic user interactions both within and across distinct periods. The user behavior sequence is compressed to different scales with the dynamic interest period scaling module in each mixer layer. The dynamic interest period scaling module is implemented with convolution layers to learn patterns over different periods of user sequence and identify local dependencies between adjacent user behaviors. The output is calculated as 
$O_i=CNN(Z_i)$, where $O_i\in R^{L',D}$ is the output of the dynamic interest period scaling module, $Z_i\in R^{L,D}$ is user sequence embedding processed by the adaptive fusion module, and $L'$ as well as $L$ follow $L'<L$ inequality. 
The bottom mixer layer maintains more details of user behaviors, while the top one mines long-term interests. Multi-scale user behavior sequence embedding extracts both low-level detailed user interests and high-level long-term representations. These cross-period user interests are then fed to the prediction layer.

\section{Experiments}
To show the effectiveness of Pyramid Mixer and reveal reasons, we conduct experiments and seek answers to the following key research questions:
\begin{itemize}
\item \textbf{RQ1:} Can our proposed method outperform state-of-the-art baselines in sequential recommendation task?
\item \textbf{RQ2:} What are the effects of key components of Pyramid Mixer?
\item \textbf{RQ3:} How does Pyramid Mixer perform in the online recommendation system?
\end{itemize}
\subsection{Experimental Settings}
\textbf{Datasets.} 
We take \textbf{MovieLens-100k}~\cite{harper2015movielens} and \textbf{MovieLens-1M} as our datasets. MovieLens is a widely used movie rating dataset in recommender systems. We also use the "Beauty" category of Amazon online reviews~\cite{mcauley2015image} in our experiments. We follow the data preprocessing method in previous studies~\cite{zhao2021recbole,mlp4rec}. Besides, we also conduct experiments on our large-scale industrial datasets. 

\textbf{Baselines.} We compare our method with the following baselines in the sequential recommendation research field. PopRec recommends the most popular items. BPR~\cite{rendle2012bpr} and FPMC~\cite{rendle2010factorizing} are classic sequential recommendation baselines. GRU4Rec~\cite{hidasi2015session}, BERT4Rec~\cite{sun2019bert4rec}, and SASRec~\cite{kang2018self} are also fundamental studies for the sequential recommendation task. MLP-Mixer~\cite{Mlp-mixer} propose an all-MLP architecture. FMLP-Rec~\cite{fmlp-rec} and MLP4Rec~\cite{mlp4rec} use all-MLP architecture for sequential recommendation purposes.


\textbf{Offline Evaluation.} We use Hit Ratio~(HR), Normalized Discounted Cumulative Gain~(NDCG), and Mean Reciprocal Rank~(MRR) as metrics on public datasets. 
For the industrial offline ablation study, we take AUC and UAUC~\cite{zhou2018deep} as our evaluation metrics of ranking performance, and we use calculation FLOPs and parameter amounts as model efficiency evaluation metrics.

\textbf{Online Evaluation.} Since calculating Daily Active Users~(DAU) directly in A/B tests is challenging, we adopt Active Days and Active Hours, defined as the number of days and hours a user is active over recent months, as surrogate metrics to evaluate improvements in user experience~\cite{yan2024trinity}. Additionally, we measure user engagement through time-based indicators, including Stay Duration (the average time a user spends in the app) and Playtime (the time spent watching videos). To assess user interactions, we incorporate a range of metrics, including Publish, Play, Finish, Like, Dislike, Comment, Share, and Follow.

\subsection{Overall Performance~(RQ1)} 
We compare our proposed method Pyramid Mixer with the state-of-the-art baselines on three public datasets, and the results are shown in Table~\ref{offline_exp_result}. We can observe that Pyramid Mixer outperforms these baseline models on all of these three datasets. On larger datasets such as Beauty, our model performs even better and demonstrates strong scalability.
\begin{table*}[t]\small\vspace{-1\baselineskip}
\caption{Offline Overall Performance Comparison Results.}\vspace{-0.5\baselineskip}
\label{offline_exp_result}
\centering
\scalebox{1}{
\begin{tabular}{c|lll|lll|lll}\toprule
Method        & \multicolumn{3}{c|}{MovieLens-100K}                                                      & \multicolumn{3}{c|}{MovieLens-1M}                                                        & \multicolumn{3}{c}{Beauty}                                                              \\ \midrule
Metrics       & \multicolumn{1}{c}{MRR@10}    & \multicolumn{1}{c}{NDCG@10}   & \multicolumn{1}{c|}{HR@10} & \multicolumn{1}{c}{MRR@10}    & \multicolumn{1}{c}{NDCG@10}   & \multicolumn{1}{c|}{HR@10} & \multicolumn{1}{c}{MRR@10}    & \multicolumn{1}{c}{NDCG@10}   & \multicolumn{1}{c}{HR@10} \\ \midrule
PopRec        & \multicolumn{1}{c}{0.1496} & \multicolumn{1}{c}{0.0783} & 0.4044                       & \multicolumn{1}{c}{0.3904} & \multicolumn{1}{c}{0.4547} & 0.6599                       & \multicolumn{1}{c}{0.0305} & \multicolumn{1}{c}{0.0443} & 0.0954                       \\ 
BPR           & \multicolumn{1}{c}{0.1479} & \multicolumn{1}{c}{0.1905} & 0.3474                       & \multicolumn{1}{c}{0.3684} & \multicolumn{1}{c}{0.4425} & 0.6781                       & \multicolumn{1}{c}{0.1310} & \multicolumn{1}{c}{0.1725} & 0.3055                       \\ 
FPMC          & \multicolumn{1}{c}{0.1220} & \multicolumn{1}{c}{0.1813} & 0.3810                       & \multicolumn{1}{c}{0.3413} & \multicolumn{1}{c}{0.4165} & 0.6568                       & \multicolumn{1}{c}{0.1431} & \multicolumn{1}{c}{0.1838} & 0.3165                       \\ 
GRU4Rec       & \multicolumn{1}{c}{0.1860} & \multicolumn{1}{c}{0.2550} & 0.4758                       & \multicolumn{1}{c}{0.4015} & \multicolumn{1}{c}{0.4658} & 0.6709                       & \multicolumn{1}{c}{0.1632} & \multicolumn{1}{c}{0.2050} & 0.3417                       \\ 
SASRec        & \multicolumn{1}{c}{0.1901} & \multicolumn{1}{c}{0.2612} & 0.4920                       & \multicolumn{1}{c}{0.4100} & \multicolumn{1}{c}{0.4720} & 0.6697                       & \multicolumn{1}{c}{0.2009} & \multicolumn{1}{c}{0.2447} & 0.3874                       \\ 
BERT4Rec      & \multicolumn{1}{c}{0.1819} & \multicolumn{1}{c}{0.2568} & 0.5061                       & \multicolumn{1}{c}{0.3938} & \multicolumn{1}{c}{0.4572} & 0.6593                       & \multicolumn{1}{c}{0.1313} & \multicolumn{1}{c}{0.1738} & 0.3135                       \\ 
FMLP          & \multicolumn{1}{c}{0.1809} & \multicolumn{1}{c}{0.2574} & 0.5070                       & \multicolumn{1}{c}{0.3932} & \multicolumn{1}{c}{0.4570} & 0.6606                       & \multicolumn{1}{c}{0.1449} & \multicolumn{1}{c}{0.1891} & 0.3336                       \\ 
MLP-Mixer     & \multicolumn{1}{c}{0.1987} & \multicolumn{1}{c}{0.2671} & 0.4920                       & \multicolumn{1}{c}{0.4110} & \multicolumn{1}{c}{0.4770} & 0.6871                       & \multicolumn{1}{c}{0.2089} & \multicolumn{1}{c}{0.2556} & 0.4065                       \\ 
MLP4Rec       & \multicolumn{1}{c}{0.2027} & \multicolumn{1}{c}{0.2747} & 0.5118                       & \multicolumn{1}{c}{0.4312} & \multicolumn{1}{c}{0.4938} & 0.6919                       & \multicolumn{1}{c}{0.2139} & \multicolumn{1}{c}{0.2654} & 0.4326                       \\ \midrule
Pyramid Mixer & \multicolumn{1}{c}{0.2043} & \multicolumn{1}{c}{0.2768} & 0.5197                       & \multicolumn{1}{c}{0.4432} & \multicolumn{1}{c}{0.4968} & 0.6993                       & \multicolumn{1}{c}{0.2293} & \multicolumn{1}{c}{0.2668} & 0.4412                       \\ \bottomrule
\end{tabular}
}
\end{table*}

\begin{table}[t]
\vspace{-0.5\baselineskip}
\caption{Ablation of Low-rank Decomposition.}\vspace{-0.5\baselineskip}
\label{Ablation of Low-rank Decomposition}
\begin{tabular}{l|lll}\toprule
Method &  AUC & FLOPs   & Parameters \\ \midrule
\begin{tabular}[c]{@{}l@{}}Pyramid Mixer\\ w/o Low rank\end{tabular} & +0.10\% & +18.0\% & +5.6\% \\ \hline
Pyramid Mixer              & +0.10\% & \textbf{+9.6\%}  & \textbf{+4.6\%} \\ \bottomrule
\end{tabular}
\end{table}

\subsection{Ablation Study~(RQ2)} 
The Pyramid Mixer model proposes three different user interest crossing modules to mine user preferences comprehensively and also utilizes a low-rank decomposition trick for efficiency improvement. In this section, we conduct experiments on our industry dataset to validate the effectiveness of each sub-module proposed in the Pyramid Mixer model. 

\begin{figure}[t]\small
  \centering\vspace{-1\baselineskip}
  \includegraphics[width=1\linewidth]{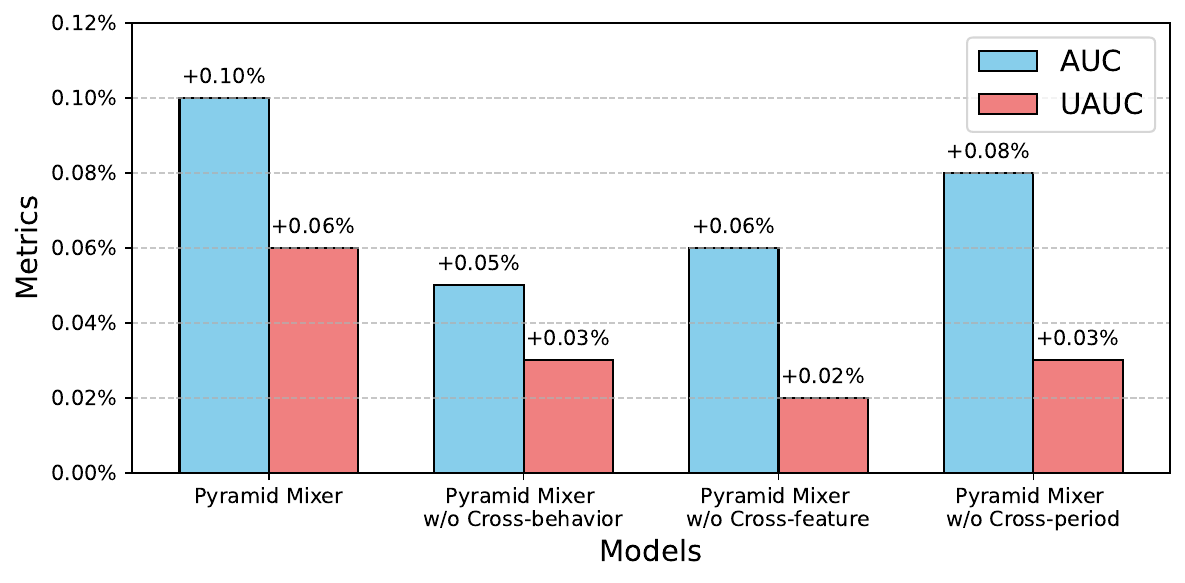}
  \vspace{-2\baselineskip}
  \caption{Ablation of User Sequence Crossing Modules.}\vspace{-1\baselineskip}
  \label{Ablation of User Behavior Crossing Methods}
\end{figure}

In Figure~\ref{Ablation of User Behavior Crossing Methods}, we compare the performance of the Pyramid Mixer without cross-behavior, without cross-feature, without cross-period, and the complete model. We only keep the Cross-behavior Mixer Module or Cross-feature Mixer Module and remove the other one to examine the performance of the Pyramid Mixer without cross-behavior and Pyramid Mixer without cross-feature. And Pyramid Mixer without cross-period removes multi-layer stacking to disable the cross-period modeling ability. The complete model leads to +0.1\% Finish AUC incremental, which is regarded as significant for the ranking model of the Douyin app. After removing cross-behavior, cross-feature, or cross-period modules, the AUC improvement bears some loss. The Finish AUC and UAUC results demonstrate that every crossing method helps Pyramid Mixer understand user interests better.

Table~\ref{Ablation of Low-rank Decomposition} shows that low-rank decomposition improves efficiency and maintains effectiveness. Low-rank decomposition leads to the same AUC results indicating that it does not undermine the overall performance of the model. However, the low-rank decomposition trick significantly improves the training efficiency by halving the model FLOPs incremental and reducing parameters to be added.

\subsection{Online Experiments~(RQ3)} To evaluate the performance of the proposed method in a real-world recommendation system, we integrated it into our ranking model to serve our largest recommendation scenario and conducted an online A/B test.

The core-metric results, presented in Table~\ref{Online Core-Metric}, reveal significant improvements in user active days, active hours, stay duration, and playtime, indicating that our model effectively improves user experience and then encourages users to spend more time on the app. Additionally, the user interaction metrics summarized in Table~\ref{Online Interaction} demonstrate that the proposed model enhances user engagement within the community.

\begin{table}[t]
\small
\caption{Online Experiment Core-Metric Results.}\vspace{-0.5\baselineskip}
\label{Online Core-Metric}
\begin{tabular}{l|llllllllllll}
\toprule
Metric     & Active Days & Active Hours & Stay Duration & Playtime    \\ \midrule
A/B result & +0.0403\% & +0.0476\%  & +0.0853\%      & +0.1106\% \\\bottomrule
\end{tabular}\vspace{-0.5\baselineskip}
\end{table}

\begin{table}[t]
\caption{Online Experiment Interaction Results.}\vspace{-0.5\baselineskip}
\label{Online Interaction}
\begin{tabular}{l|lllllllll}
\toprule

Metric     & Publish  & Play      & Finish  & Like        \\ \midrule
A/B result & +0.093\% & +0.0538\% & +0.1801\% & +0.1761\% \\\bottomrule

Metric     & Dislike   & Comment   & Share     & Follow    \\ \midrule
A/B result & -0.0145\% & +0.2074\% & +0.2284\% & +0.4876\% \\\bottomrule

\end{tabular}
\end{table}

\section{Conclusion}
In this paper, we introduce the Pyramid Mixer model for sequential recommendation tasks. With a multi-dimensional mixer network, the proposed model effectively captures comprehensive cross-behavior and cross-feature user interests. Additionally, the hierarchical stacking of mixer layers in a pyramid structure facilitates the exploration of cross-period user preferences. Extensive offline and online experiments conducted on the industrial platform further validate the effectiveness and efficiency of the Pyramid Mixer model.
\bibliographystyle{ACM-Reference-Format}
\balance
\bibliography{reference}
\appendix

\end{document}